\documentclass{ws-ijmpe}
\usepackage{graphicx}
\usepackage{epsfig}
\usepackage{psfrag}
\usepackage[english]{babel}
\usepackage{amsmath}
\usepackage{verbatim}
\usepackage{mathrsfs}
\usepackage{amsfonts}
\usepackage{amssymb}
\usepackage{hyperref}
\usepackage{float}
\usepackage{color}
\usepackage{cancel}

\DeclareMathOperator{\tr}{tr}

\begin{document}

\markboth{K. Fushimi, M. E. Mosquera, O. Civitarese}{MSSM
WIMPs-nucleon cross section for E$_\chi <$ 500 GeV}

\catchline{}{}{}{}{}
%%%%%%%%%%%%%%%%%%%%%%%%%%%%%%%%%%%%%%%%%%%%%%%%%%%%%%%%%%%%%%%%%%%%

\title{MSSM WIMPs-nucleon cross section for E$_\chi <$ 500 GeV}

\author{K. Fushimi}

\address{Facultad de Ciencias Astron\'omicas y Geof\'{\i}sicas, University of La Plata. Paseo del Bosque S/N\\
1900, La Plata, Argentina.\\
keiko.fushimi@fcaglp.unlp.edu.ar}

\author{M. E. Mosquera}

\address{Dept. of Physics, University of La Plata, c.c.~67\\
Facultad de Ciencias Astron\'omicas y Geof\'{\i}sicas, University of La Plata. Paseo del Bosque S/N\\
 1900, La Plata, Argentina\\
mmosquera@fcaglp.unlp.edu.ar
}

\author{O. Civitarese}

\address{Dept. of Physics, University of La Plata, c.c.~67\\
 1900, La Plata, Argentina\\
osvaldo.civitarese@fisica.unlp.edu.ar}

\maketitle

\begin{history}
\received{Day Month Year}
\revised{Day Month Year}
%\accepted{Day Month Year}
%\comby{(xxxxxxxxxx)}
\end{history}
\begin{abstract}
Among dark-matter candidates are the WIMPs (Weekly Interacting Massive Particles). Low-threshold detectors could directly detect dark-matter by measuring the energy deposited by the particles. In this work we examine the cross section for the elastic scattering of WIMPs on nucleons, in the spin-dependent and spin-independent channels. WIMPs are taken as neutralinos in the context of the minimal super-symmetric extension of the standard model (MSSM). The dependence of the results with the adopted MSSM parameters is discussed.
\end{abstract}

\maketitle

%-------------------------------------------------
%  INTRO
%-------------------------------------------------
\section{Introduction}
\label{sec:intro}

The Universe is mostly compossed by dark matter (DM), which does not
shine but interacts gravitationally with baryonic matter. The first
evidence about the existence of such kind of matter were proposed by
Zwicky in the 30s and in the 70s  by Rubin et al.
\cite{Zwicky:1933,Rubin:1970}. Measurements of the Cosmic Microwave
Background (CMB) and of the gravitational-lensing effects have
produced new evidences about the existence of dark matter
\cite{Gelmini:2017,Roszkowski:2018}.

Dark matter may be produced by thermal processes, such as the collision of plasmas, and
non-thermal processes, such as particle decays \cite{majumdar:2014}. The mass of the dark matter particles is unknown. It can be as large as $10^{13}$ GeV. There exist several candidates to dark matter, e. g. WIMPZILLA, MACHO, axions, sterile neutrinos, WIMPs.

The WIMPZILLA is the heaviest candidate of all and it could be produced through non-thermal processes after the inflation epoch \cite{Chung:2001}. These particles could decay in ultra-energetic cosmic rays \cite{Kolb:1999,Alcantara:2019}. The MACHOs (massive astrophysical halo objects) are compact objects of baryonic matter. Some examples are brown-dwarf (mass of the order of $0.08 \, M_{\odot}$), Jupiter's kind objects (mass of the order of $0.001 \, M_{\odot}$), white-dwarf, planets and primordial black holes \cite{Undagoitia:2015}. These objects can be detected through the gravitational micro-lensing effect \cite{Paczynski:1986}. MACHOs may represent  $20\%$ of the dark matter in the galaxy. Models that considered MACHOs as dominant components of dark matter are rule out \cite{Alcock:2000}. The baryonic component of dark
matter can not explain the results of Planck \cite{Planck:2018}, therefore there should be non-baryonic dark matter. The axions are bosons introduced to solve the CP-problem \cite{majumdar:2014} with masses of the order of $10^{-9}$ eV \cite{Perez:2020}. The interaction between axions and photons through the Primakoff effect gives photons that could be detected \cite{Freese:2017}. The sterile neutrino were proposed as warm dark matter candidate in Ref. \cite{Dodelson:1994}. The mass of these particles should be lower than $1$ keV \cite{Boyarsky:2019}. There are several experimental efforts to set constrains on the mass of this neutrino and on the active-sterile mixing angle \cite{Campos:2016,Divari:2017}.

Weakly interacting massive particles (WIMPs) are the most probably
dark-matter candidates. These particles interacts through weak
interactions with matter \cite{Gelmini:2017}, their mass could be in
the range 1 GeV - 10 TeV \cite{Freese:2012}, however, this
mass-range can be reduced to 3 GeV - 20 GeV \cite{davis:2015}. Some
of these candidates are the heavy photon (little Higgs theories),
the Kaluza-Klein photon (multidimensional theories) and the
neutralino (super-symmetric theories) \cite{Scopel:2008}. Since the
evidence suggests that dark matter is massive, stable, cold, and
electrically neutral, neutralinos are preferred other than more
exotic candidates. Recent measurements of the excess in the electron
recoil from XENON1T experiment \cite{xenon}, may be taken as
indication about the existence of  two-components-exothermic dark
matter \cite{lee}. The theoretical description implies the use of
effective Lagrangians where the interactions of dark matter
particles with neutral massive bosons is explicitly taken into
account, as done in \cite{lee}. There exist two different methods to
detect WIMPs: direct and indirect detections. The former aims at
measuring the energy deposited by a WIMP when it interacts with the
detector \cite{Gelmini:2016,Goodman:1985,Drukier:1986}, and the
latter looks for annihilation or decay products of WIMPs such as
gamma rays, neutrinos, and cosmic rays \cite{Conrad:2014}. In this
work, we shall focus on the calculation of cross sections relevant
for direct detection experiments.

Because of the extremely low signal-to-noise rates, direct-detection
experiments need to be performed in low-background conditions,
therefore, they are located in underground laboratories
\cite{Bernabei:2018,Amare:2019,cdex:2018,Abdelhameed:2019,CDMS:2016,XENON1T:2018,Antonello:2018,Santos:2013,Adhikari:2019,DarkSide-50:2018,Amole:2019,Ajaj:2019}.
A modulation of the signal by the movement of the Earth in its
orbit, phenomenon called annual modulation, is also expected to be
measured \cite{Bernabei:2018} , as well as the diurnal modulation
which is produced by the Earth movement around its axis.

The currently running detectors are located in the northern hemisphere, however there exist projects to settled detectors in the southern hemisphere, like in the planned ANDES laboratory in Agua Negra (Argentina) \cite{ANDES_LAB} and the SABRE experiment (Australia, not yet operational) \cite{Antonello:2018}. The experimental data, when available, will allow for the comparison between nothern and southern located
detectors, hopefully.

In this work we have considered the lightest neutralino as a possible candidate for  dark matter. We have computed the scattering cross section between the neutralino and protons (neutrons) as a function of different parameters of the MSSM (minimal super-symmetric extension of the standard model). Then, we have compared the theoretical results with the available experimental data.

The paper is organized as follows. In section \ref{formalismo} we provide a brief description of neutralinos's properties and the formalism needed to compute the neutralino-nucleon cross section. In section \ref{resultados} we show and discussed the results of the calculations. The conclusions are drawn in section \ref{conclusion}.

%-------------------------------------------------
%  NEUTRALINO
%-------------------------------------------------
\section{Formalism}
\label{formalismo}

In the framework of the MSSM  models, each elementary particle has a super-partner with a spin that differs by a half-integer. The mass-matrix of the sector of neutral fermions can be written, in terms of the bino, wino and higgsino mass parameters ($M$, $M'$ and $\mu$ respectively) as \cite{Elkheishen:1992,bertone04}
\begin{eqnarray}\label{m-matrix}
Y&=&\left(
\begin{array}{cccc}
 M' & 0 & -M_Z c_{\beta} s_{W} & M_Z s_{\beta} s_W\\
 0 & M & M_Z c_{\beta} c_W & - M_Z s_{\beta} c_W\\
-M_Z c_{\beta} s_W & M_Z c_{\beta} c_W & 0 & -\mu\\
M_Z s_{\beta}s_W & - M_Z s_{\beta} c_W & -\mu & 0\\
\end{array} \right)\, , \nonumber
\end{eqnarray}
where $c_{\beta}$ $\left(s_{\beta}\right)$ stands for $\cos{\beta}$ $\left(\sin{\beta}\right)$ with $\tan{\beta}=v_1/v_2$ \cite{Murakami:2001,Ellis:2000,Cerdeno:2001}, $M_Z$ is the mass of the $Z$-boson, $\theta_W$ is the Weinberg angle and $c_W$ $\left(s_W\right)$ is $\cos{\theta_W}$ $\left(\sin{\theta_W}\right)$. In the Grand-Unified-Theory (GUT), the parameters $M$ and $M'$ are related by $M' = \frac{5}{3} M \tan^2{\theta_W}$.

The lightest neutralino state may be written as a linear combination of binos ($\tilde{B}$), winos ($\tilde{W}_3$) and higgsinos ($\tilde{H}^0_1$, $\tilde{H}^0_2$)
\begin{eqnarray}\label{nchi}
\chi_1^0&=&Z_{11}\tilde{B}+Z_{12}\tilde{W_3}+Z_{13}\tilde{H}^0_1+Z_{14}\tilde{H}^0_2 \label{chi}\, .
\end{eqnarray}
where $Z_{11},\, Z_{12},\, Z_{13},\, Z_{14}$ are the components of the eigenvector related to the lowest positive eigenvalue of the mass matrix (\ref{m-matrix}) (see Appendix \ref{diagonalizacion} for details).

\begin{figure}[h!]
\epsfig{file=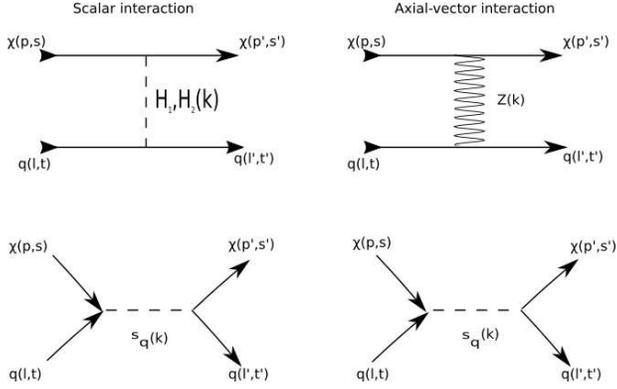, width=230pt, angle=0}
\caption{Contributions to the axial-vector and scalar interactions
of the Lagrangian (\ref{lagrangian}) at tree level.}
\label{diagrama2}
\end{figure}

The effective Lagrangian density describing the neutralino-quark elastic-scattering in the MSSM is given by \cite{Engel:1992}:
\begin{eqnarray}\label{lagrangian}
\mathcal{L}_{eff}&=&\frac{g^2}{2M_W^2}\sum_q\left(\bar{\chi}\gamma^{\mu}\gamma_5\chi\bar{\psi_q}\gamma_{\mu}A_q\gamma_5\psi_q \right. \nonumber \\
&&\left. \hskip 1.5cm +\bar{\chi}\chi S_q \frac{m_q}{M_W}\bar{\psi_q}\psi_q\right)\, .
\end{eqnarray}
In Equation (\ref{lagrangian}) $\psi_q$ represents the quark field, $\chi$ is the dark matter field, g is the SU(2) coupling constant and $M_W$ stands for the mass of the W boson. The axial-vector ($\gamma_{\mu}\gamma_{5}$) and scalar terms of it are shown in Fig. \ref{diagrama2}. The coupling constants, $A_q$ and $ S_q$ can be written as \cite{Engel:1992}
\begin{eqnarray}
A_q&=&-\frac{M^2_W}{M^2_{\tilde{q}}}\left[\left(T_{3q}Z_{12}-\tan{\theta_W}(T_{3q}-e_q)Z_{11}\right)^2\right. \nonumber \\
&&\left. \hskip 1.5cm +\tan^2{\theta_W} e_q^2Z_{11}^2+ \frac{m_q^2d_q^2}{2M_W^2} \right]\nonumber \\
&&+\frac{T_{3q}}{2} (Z^2_{13}-Z^2_{14}) ,\nonumber\\
S_q&=&\frac{Z_{12}-\tan{\theta_W}Z_{11}}{2}  \nonumber \\ &&\times\left(\frac{M^2_W}{M^2_{H^0_2}}g_{H^0_2}k_q^{(2)}+\frac{M^2_W}{M^2_{H^0_1}}g_{H^0_1}k_q^{(1)}+\frac{M^2_W \epsilon d_q}{M^2_{\tilde{q}}}\right),\nonumber
\end{eqnarray}
where $M_{\tilde{q}}$ and $M_{H^0_i}$ are the squark and higgsino
masses respectively \cite{Djouadi:2008,Tanabashi:2018}, $T_{3q}$,
$e_q$ and $m_q$ are the quark weak isospin, charge and mass
respectively, $\epsilon$ is the sign of the lightest-neutralino mass
eigenvalue \cite{Engel:1992}. The corresponding values of $d_q$ and
$k_q^{(i)}$ for the up-type and down-type quark are given in Table
\ref{quark} \cite{Djouadi:2008}. The constants for the higgsinos are
\begin{eqnarray}
g_{H^0_1}&=&-Z_{13}\cos{\alpha}+Z_{14}\sin{\alpha}\, , \nonumber \\
g_{H^0_2}&=&Z_{13}\sin{\alpha}+Z_{14}\cos{\alpha}\, , \nonumber \\
\alpha&=&0.5 \arctan\left(\frac{m_{ps}^2+M_Z^2}{m_{ps}^2-M_Z^2}\tan \left(2\beta\right)\right)\, . \nonumber
\end{eqnarray}
In the previous equations $m_{ps}$ is the mass of the Higgs
pseudo-scalar \cite{Djouadi:2008}.

\begin{table}[h!]
\begin{center}
\begin{tabular}{|c|c|c|c|}
\hline
quark & $d_q$ & $k_q^{(1)}$ & $k_q^{(2)}$ \\ \hline
up & $\frac{Z_{14}}{\sin{\beta}}$ & $\frac{\sin{\alpha}}{\sin{\beta}}$ & $\frac{\cos{\alpha}}{\sin{\beta}}$ \\ \hline
down & $\frac{-Z_{13}}{\cos{\beta}}$ & $\frac{\cos{\alpha}}{\cos{\beta}}$ & $\frac{-\sin{\alpha}}{\cos{\beta}}$ \\ \hline
\end{tabular}
\end{center}
\caption{Values of $d_q$ and $k_q^{(i)}$ for the up-type and
down-type quark adopted for the calculations}\label{quark}
\end{table}

The cross section of the neutralino quark scattering can be computed as
\begin{eqnarray}
\sigma %\left (q\chi\longrightarrow q\chi \right)
&=& V \int \frac{ \textrm{d}^3{\bf p'}}{(2\pi)^3} V \int \frac{ \textrm{d}^3{\bf l'}} {(2\pi)^3} \sum_{s,t,s',t'} \frac{W\left(q \chi \longrightarrow q \chi\right)}{J_{\chi}V^{-1}} \, . \label{cs}
\end{eqnarray}
The integration is performed in the space of outgoing momenta ${\bf p'},\, {\bf l'}$ of the neutralino and quark, respectively, and the sum is computed over the incoming state spins $s,\, t$ and the outgoing state spins $s', \, t'$. The transition rate per unit volume is
\begin{eqnarray}
W(q\chi\longrightarrow q\chi)&=& \frac{1}{VT}\left|S(q\chi\longrightarrow q\chi)\right|^2 \nonumber \\
&=& \frac{1}{VT}\left|-i\int{ \textrm{d}t \textrm{H}_{\textrm{int}}(q{\chi}\longrightarrow q\chi)}\right|^2\, , \nonumber
\end{eqnarray}
where $\textrm{H}_{\textrm{int}}$ is the Hamiltonian and the flux of incoming neutralinos is
\begin{eqnarray}
J_{\chi}&=& \frac{\sqrt{(p_{\mu}l^{\mu})^2-m_{\chi}^2m_q^2}}{p^0l^0V}\, . \nonumber
\end{eqnarray}

In the previous equation $m_{\chi}$ and $p$ are the mass and the incoming momentum of the neutralino, $m_q$ and $l$ are the mass and the incoming momentum of the quark, respectively.

The Eq. (\ref{cs}) is reduced to
\begin{eqnarray}
\sigma%(q\chi\longrightarrow q\chi)
&=&\sum_{s',t'}\frac{1}{8 s\sqrt{\beta_-\beta_+} (2\pi)^2}\nonumber \\
&&\hskip 1cm \times \int\frac{\textrm{d}^3{\bf p'}}{p'^0} \frac{\textrm{d}^3{\bf l'}}{l'^0} \delta^4(p'-p+l'-l) \vert M\vert^2\, , \label{sigma}
\end{eqnarray}
where we have defined $\beta_{\pm}=\left(1-\frac{(m_{\chi}\pm m_q)^2}{q^2}\right)$ and $s=(p+l)^2$.

For each mediator the expression of the factor $|M|^2$ for the axial-vector $\left(A-V\right)$ and scalar $\left(S\right)$
terms of the interactions is given by
\begin{eqnarray}
M^{A-V}_Z&=& \mathcal{C}^{A-V}_{Z}\Big[\bar{\chi} \gamma^{\mu}\gamma_5\chi\Big] \Big[\bar{\psi_q}\gamma_{\mu}\gamma_5\psi_q\Big]\, , \nonumber\\
M^{A-V}_{\tilde{q}}&=& \mathcal{C}^{A-V}_{\tilde{q}} \Big[\bar{\chi}\gamma^{\mu}\gamma_5\chi\Big] \Big[\bar{\psi_q}\gamma_{\mu}\gamma_5\psi_q\Big] \, ,\nonumber\\
M^{S}_{H}&=& \mathcal{C}^{S}_{H}\Big[\bar{\chi}\chi\Big] \Big[\bar{\psi_q}\psi_q\Big]\, , \nonumber\\
M^{S}_{\tilde{q}}&=& \mathcal{C}^{S}_{\tilde{q}} \Big[\bar{\chi}\chi\Big] \Big[\bar{\psi_q}\psi_q\Big]\, ,\nonumber
\end{eqnarray}
with
\begin{eqnarray}
\mathcal{C}^{A-V}_{Z}&=&\frac{g^2}{4\cos^2{\theta_W}(k^2-M_Z^2)} T_{3q}(Z^2_{13}-Z^2_{14})\, , \nonumber\\
\mathcal{C}^{A-V}_{\tilde{q}}&=&\frac{g^2}{2(s-M^2_{\tilde{q}})}\left[\left(T_{3q}Z_{12}-\tan{\theta_W}(T_{3q}-e_q)Z_{11}\right)^2\right. \nonumber \\
&&\left. \hskip 2cm +\tan^2{\theta_W}e_q^2 Z_{11}^2 +\frac{2m_q^2d_q^2}{4M_W^2}\right]\, , \nonumber\\
\mathcal{C}^{S}_{H}&=&\frac{g^2 m_q}{4M_W} \left(Z_{12}- \tan{\theta_W}Z_{11}\right) \nonumber \\
&&\times\left(\frac{g_{H^0_2}k_q^{(2)}}{(k^2-M^2_{H^0_2})} +\frac{g_{H^0_1}k_q^{(1)}}{(k^2-M^2_{H^0_1})}\right)\, , \nonumber\\
\mathcal{C}^{S}_{\tilde{q}}&=&\frac{g^2 m_q}{4M_W} \left(Z_{12}-\tan{\theta_W}Z_{11}\right)\frac{\epsilon d_q}{(s-M^2_{\tilde{q}})}\, . \nonumber
\end{eqnarray}
In the previous definitions $k$ is the momentum of the mediator. In the following we are going to analyse the axial-vector and scalar channel channels separately.

\begin{itemize}
\item Axial-vector channel

The matrix elements for the axial-vector interaction is written as
\begin{eqnarray}
M_{A-V}&=&\left(\mathcal{C}^{A-V}_{Z}+\mathcal{C}^{A-V}_{\tilde{q}}\right)\Big[\bar{\chi} \gamma^{\mu}\gamma^5 {\chi}\Big] \Big[\bar{\psi}_{q}\gamma_{\mu}\gamma_5 {\psi}_{q}\Big] \, , \nonumber
\end{eqnarray}
After some algebra (see appendix \ref{ap-AV} for detail), we have obtained
\begin{eqnarray}
\sum_{s,s',t,t'}\vert M\vert_{A-V}^2 &=& 32\left(\mathcal{C}^{A-V}_{Z}+\mathcal{C}^{A-V}_{\tilde{q}}\right)^2\nonumber\\
&&\times \left[(p \cdot l)(p'\cdot l')+(p\cdot l')(p'\cdot l)  \right. \nonumber\\
&&\left. \hskip 0.5cm+m_{\chi}^2(l\cdot l') +m_q^2(p\cdot p') +2m_q^2m_{\chi}^2\right]\nonumber
\end{eqnarray}

Using Eq. (\eqref{sigma}) the cross section reads
\begin{eqnarray}
\sigma_{A-V}&=&\frac{32}{8 s\sqrt{\beta_-\beta_+} (2\pi)^2}\int\frac{\textrm{d}^3{\bf p'}}{p'^0}\frac{\textrm{d}^3{\bf l'}}{l'^0} \delta^4(p'-p+l'-l)\nonumber\\
&&\hskip 3cm \times \left(\mathcal{C}^{A-V}_{Z}+\mathcal{C}^{A-V}_{\tilde{q}}\right)^2 \nonumber\\
&&\hskip 3cm \times \left[(p \cdot l)(p'\cdot l')+(p\cdot l')(p'\cdot l)\right.\nonumber\\
&&\left. \hskip 3.5cm +m_{\chi}^2(l\cdot l')+m_q^2(p\cdot  p')\right.\nonumber\\
&&\left. \hskip 3.5cm +2m_q^2m_{\chi}^2\right]\nonumber
\end{eqnarray}
and after performing the integration it yields
\begin{eqnarray}
\sigma_{A-V}&=&\frac{1}{\pi}\left(\frac{\left(\mathcal{\tilde{C}}^{A-V}_{Z}\right)}{\lambda_q^{Z}} + \frac{\left(\mathcal{\tilde{C}}^{A-V}_{\tilde{q}}\right)} {\left(s-M_{\tilde{q}}^2\right)}\right)^2 \nonumber\\
&&\times\left\{\frac{s}{6}\left[1-\frac{m_{\chi}^2+m_{q}^2}{s}+ \frac{4 m_q^2m_{\chi}^2}{s^2} \right. \right.\nonumber\\
&&\left. \left. \hskip 1.5cm
-\left(\frac{m_{\chi}^2-m_{q}^2}{s}\right)^2 \frac{m_{\chi}^2+m_{q}^2}{s} \right. \right.\nonumber\\
&&\left. \left. \hskip 1.5cm +\left(\frac{m_{\chi}^2-m_{q}^2}{s}\right)^4\right] \right. \nonumber\\
&&\left. \hskip 0.5 cm+\frac{s}{2}\left(1-\frac{m_q^2+m_{\chi}^2}{s}\right)^2\right. \label{csav}\\
&&\left. \hskip 0.5 cm+ \frac{m_{\chi}^2}{2}\left(1-\frac{m_{\chi}^2-m_{q}^2}{s}\right)^2\right. \nonumber\\
&&\left. \hskip 0.5 cm+\frac{m_{q}^2}{2} \left(1+\frac{m_{\chi}^2-m_{q}^2}{s} \right)^2+ \frac{4 m_{q}^2 m_{\chi}^2}{s}\right\}\, . \nonumber
\end{eqnarray}
where $p_0$ is the energy of the neutralino. We have defined
\begin{eqnarray}
\lambda_q^Z &=& 2m_{q}^2-M_{Z}^2-\frac{s}{2}\left(1-\frac{m_{\chi}^2-m_{q}^2}{s}\right)^2\, , \nonumber \\ \mathcal{\tilde{C}}^{A-V}_{Z}&=&\mathcal{C}^{A-V}_{Z}(k^2-M_Z^2)\, , \nonumber \\ \mathcal{\tilde{C}}^{A-V}_{\tilde{q}}&=&\mathcal{C}^{A-V}_{\tilde{q}}(s-M_{\tilde{q}}^2)\, . \nonumber
\end{eqnarray}

\item Scalar channel

In this case, the matrix elements is written as
\begin{eqnarray}
 M_{S}&=&\left(\mathcal{C}^{S}_{H}+\mathcal{C}^{S}_{\tilde{q}}\right)\Big[\bar{\chi}{\chi}\Big]\Big[\bar{\psi}_{q}{\psi}_{q}\Big] \nonumber
\end{eqnarray}
leading to the result (see appendix \ref{ap-SC})
\begin{eqnarray}
\sum_{s,s',t,t'}\vert M\vert_{S}^2 &=&16 \left(\mathcal{C}^{S}_{H}+\mathcal{C}^{S}_{\tilde{q}}\right)^2 \left[(p \cdot p')(l\cdot l')+m_{\chi}^2(l\cdot l')\right. \nonumber\\
&&\left. \hskip 2.5 cm+m_q^2(p\cdot p')+m_q^2m_{\chi}^2\right] \, .\nonumber
\end{eqnarray}
With this result the contribution due to the scalar channel to the cross-section is written as
\begin{eqnarray}
\sigma_{S}&=&
\frac{1}{2\pi}\left[\frac{\mathcal{\tilde{C}}^{Sc}_{H_1^0}}{\lambda_q^{H_1}}+\frac{\mathcal{\tilde{C}}^{Sc}_{H_2^0}}{\lambda_q^{H_2}}+\frac{\mathcal{\tilde{C}}^{Sc}_{\tilde{q}}}{s-M_{\tilde{q}}^2}\right]^2\nonumber\\
%&& \frac{1}{2\pi}\left[\left(\frac{\mathcal{\tilde{C}}^{S}_{H_1^0}}{\lambda_q^{H_1}} + \frac{\mathcal{\tilde{C}}^{S}_{H_2^0}}{\lambda_q^{H_2}}\right)^2 + \left(\frac{\mathcal{\tilde{C}}^{S}_{H_2^0}}{\lambda_q^{H_2}} + \frac{\mathcal{\tilde{C}}^{S}_{\tilde{q}}}{\left(s-M_{\tilde{q}}^2\right)}\right)^2\right]\nonumber\\
&&\times\left\{\frac{s}{6} \left[1-\frac{m_{\chi}^2+m_{q}^2}{s}+\frac{4 m_q^2m_{\chi}^2}{s^2} \right. \right.\nonumber\\
&&\left. \left. \hskip 1.5cm -\left(\frac{m_{\chi}^2-m_{q}^2}{s}\right)^2 \frac{m_{\chi}^2+m_{q}^2}{s}\right. \right.\nonumber\\
&&\left. \left. \hskip 1.5cm +\left(\frac{m_{\chi}^2-m_{q}^2}{s}\right)^4\right]\right.\nonumber\\
&&\left. \hskip 0.7cm+\frac{m_{\chi}^2}{2}\left(1-\frac{m_{\chi}^2-m_{q}^2}{s}\right)^2\right. \nonumber\\
&&\left. \hskip 0.5 cm +\frac{m_{q}^2}{2} \left(1+\frac{m_{\chi}^2-m_{q}^2}{s}\right)^2 +\frac{2 m_{q}^2 m_{\chi}^2}{s}\right\}\, , \label{cssc}
\end{eqnarray}
where
\begin{eqnarray}
\lambda_q^{H_{i}}&=&2m_{q}^2-M_{H_{i}}^2-\frac{s}{2}\left(1-\frac{m_{\chi}^2-m_{q}^2}{s}\right)^2\, , \nonumber \\
\mathcal{\tilde{C}}^{S}_{H_{i}}&=&\mathcal{C}^{S}_{H_{i}}(k^2-M^2_{H_{i}})\, , \nonumber \\ \mathcal{\tilde{C}}^{S}_{\tilde{q}}&=&\mathcal{C}^{S}_{\tilde{q}}(s-M_{\tilde{q}}^2) \, . \nonumber
\end{eqnarray}
\end{itemize}

To obtain the cross section of the neutralino-nucleon scattering we express the nucleon fields in terms of the quark fields. For the axial-vector interaction the cross section is written as
\begin{eqnarray}
\sigma_{A-V}^N&=& \sum_{q=u,d,s} \sigma_{A-V} \left(\Delta_q^N\right)^2\, ,
\end{eqnarray}
where $\sigma_{A-V}^N$ is the cross section of Eq. (\ref{csav}) and $\Delta_q^N$ are experimental values that describe the contribution of a quark $q$ to the spin of the nucleon $N$ (either a proton ($p$) or a neutron ($n$)) \cite{Gondolo:2004}
\begin{eqnarray}
\Delta u^p=\Delta d^n&=&0.77\, ,\nonumber\\
\Delta d^p=\Delta u^n&=&-0.40\, ,\nonumber\\
\Delta s^p=\Delta s^n&=&-0.12\, .\nonumber
\end{eqnarray}

The cross section for the scalar channel is given by \cite{Lisanti:2016}
\begin{eqnarray}
\sigma_{S}^N&=& \sum_{q=u,d,s} \sigma_{S}\left(\frac{f_{T_q}^N m_N}{m_q}\right)^2+\sum_{q=c,b,t} \sigma_{S} \left(\frac{2}{27}f_{T_G}^N \frac{m_N}{m_q}\right)^2\, . \nonumber
\end{eqnarray}
where  $\sigma_{S}^N$ is the cross section of Eq. (\ref{cssc}).
The factors $ f_{T_q}^N$ (where $N$ denotes either proton or neutron) are the fractions of the nucleon-mass accounted
for by a particular quark-flavour, defined as \cite{Gondolo:2004}
\begin{eqnarray}
\begin{array}{ccc}
f_{T_u}^{p}=0.023\, , &\hskip 1.5cm & f_{T_d}^{p}=0.034\, ,\\
f_{T_s}^{p}=0.140\, , & & f_{T_u}^{n}=0.019\, ,\\
f_{T_d}^{n}=0.041\, , & & f_{T_s}^{n}=0.140\, ,
\end{array}\nonumber
\end{eqnarray}
with $f_{T_G}^N=1-\sum_{q=u,d,s} f_{T_q}^N$.

%-------------------------------------------------
% RESULTADOS
%-------------------------------------------------
\section{Results}
\label{resultados}

In this section we present the results of our calculations of the neutralino-nucleon cross-section as a function of the different parameters of the MSSM. We have extracted the value of the masses from Ref. \cite{Tanabashi:2018} and taken the squark-mass as $1500$ GeV, the Higgs pseudo-scalar mass as $500$ GeV and $g=0.645527$. As said before, the form factors $\Delta_q^n$ and $f_{T_q}^n$ were extracted from Ref. \cite{Gondolo:2004}. For the SUSY parameter, we have used $\tan \beta=3$ \cite{Ellis:2000,Cerdeno:2001,Murakami:2001}, and varied the parameter $\mu$ and $M$ in order to obtain different values for the neutralino mass (see Table \ref{table-fig}). The parameter $M$ has a lower constraint once the value of $\mu$ is fixed, which came from the fact that the neutralino mass (see Appendix \ref{diagonalizacion}) is  positive defined. We have explored two set of values for the neutralino mass: i) $m_\chi$ greater than $1$ GeV; and ii) $m_\chi$ smaller than $1$ GeV.

\subsection{Results for $m_\chi > 1$ GeV}

\begin{itemize}

\item Axial-vector contribution to the cross section

In Fig. \ref{av-s-vs-e} we present the nucleon-neutralino cross section for the axial-vector contribution as a function of the neutralino energy. The values of the parameters used to obtain the curves shown in the Figure, are listed in Table \ref{table-fig}. The cross section for protons and neutrons are practically the same, therefore we have displayed the results for protons. In all the cases, the contribution to the cross section, when the $Z_0$-boson is the mediator, is larger than the contribution of the squark. The dependence on the mass parameter of the higgsino $\left(\mu\right)$ is noticeable. For larger values of $\mu$ the cross-section becomes smaller.
\begin{table}[h!]
\begin{center}
\begin{tabular}{|c|c|c|}
\hline
$m_\chi$ [GeV] & $\mu$ [GeV] & $M$ [GeV] \\ \hline
& $100$ & $73.57$ \\ \cline{2-3}
$5$& $500$ & $19.82$ \\ \cline{2-3}
& $1000$ & $14.01$ \\ \hline
& $100$ & $161.40$ \\ \cline{2-3}
$35$& $500$ & $76.24$ \\ \cline{2-3}
& $1000$ & $72.45$ \\ \hline
& $100$ & $511.24$ \\ \cline{2-3}
$80$& $500$ & $166.11$ \\ \cline{2-3}
& $1000$ & $162.12$ \\ \hline
\end{tabular}
\end{center}
\caption{Values of $m_\chi$, $\mu$ and $M$ used in the calculation of the cross section of Fig. \ref{av-s-vs-e} and \ref{sc-s-vs-e} for $\tan \beta=3$.}
\label{table-fig}
\end{table}

\begin{figure*}[h!]
\epsfig{file=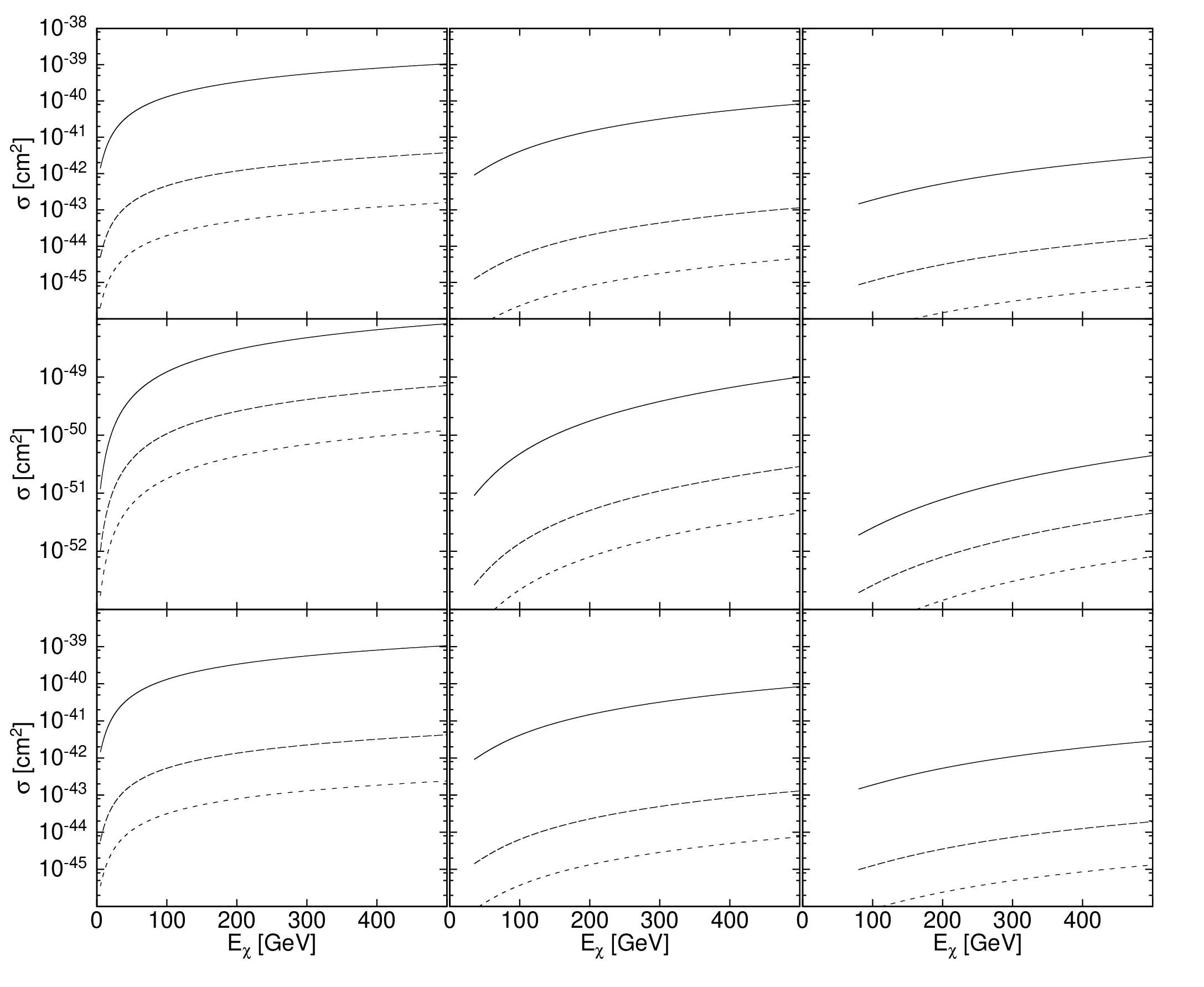, width=400pt, angle=-90}
\caption{Axial-vector contribution to the cross section as a function of the neutralino energy, $E_\chi$, for nucleon-neutralino scattering. Left column: $m_\chi=5$ GeV; middle column: $m_\chi=35$ GeV; right column: $m_\chi=80$ GeV. Top row: $Z_0$-boson contribution; middle row: squark contribution; bottom row: total axial-vector contribution. Solid line: $\mu=100$ GeV; dashed line: $\mu=500$ GeV; dotted line: $\mu=1000$ GeV. The values for $M$ are listed in Table \ref{table-fig}.}
\label{av-s-vs-e}
\end{figure*}

If the value of $\tan\beta$ is changed the cross-section does not change appreciable, since the increase of the squark contribution due to changes in this parameter (ie; for $\tan \beta=10$) does not match the contribution  due to the $Z_0$ boson as mediator. Concerning the dependence of the cross section with the quark flavour, we have seen from our results that the larger contribution comes from the strange-quark sector. We have compared our results with the limits determined by the available data \cite{behnke17,Amole:2019,xia19,suzuki19}, and found that, for the values of $\mu$ used in this work, the calculated values are in good agreement with those limits.

\item Scalar contribution to the cross section

The scalar contribution to the cross section is presented in Fig. \ref{sc-s-vs-e}, as a function of the neutralino energy. Once again, the calculated values for neutrons and protons are practically the same. The contributions for all mediators to this channel, are of the same order of magnitude. The dependence on the mass parameter of the higgsino is noticeable.
% if all the contributions are considered, however, when the mediator is an squark, for larger values of $\mu$ one obtains larger values of the cross section.
\begin{figure*}[h!]
\epsfig{file=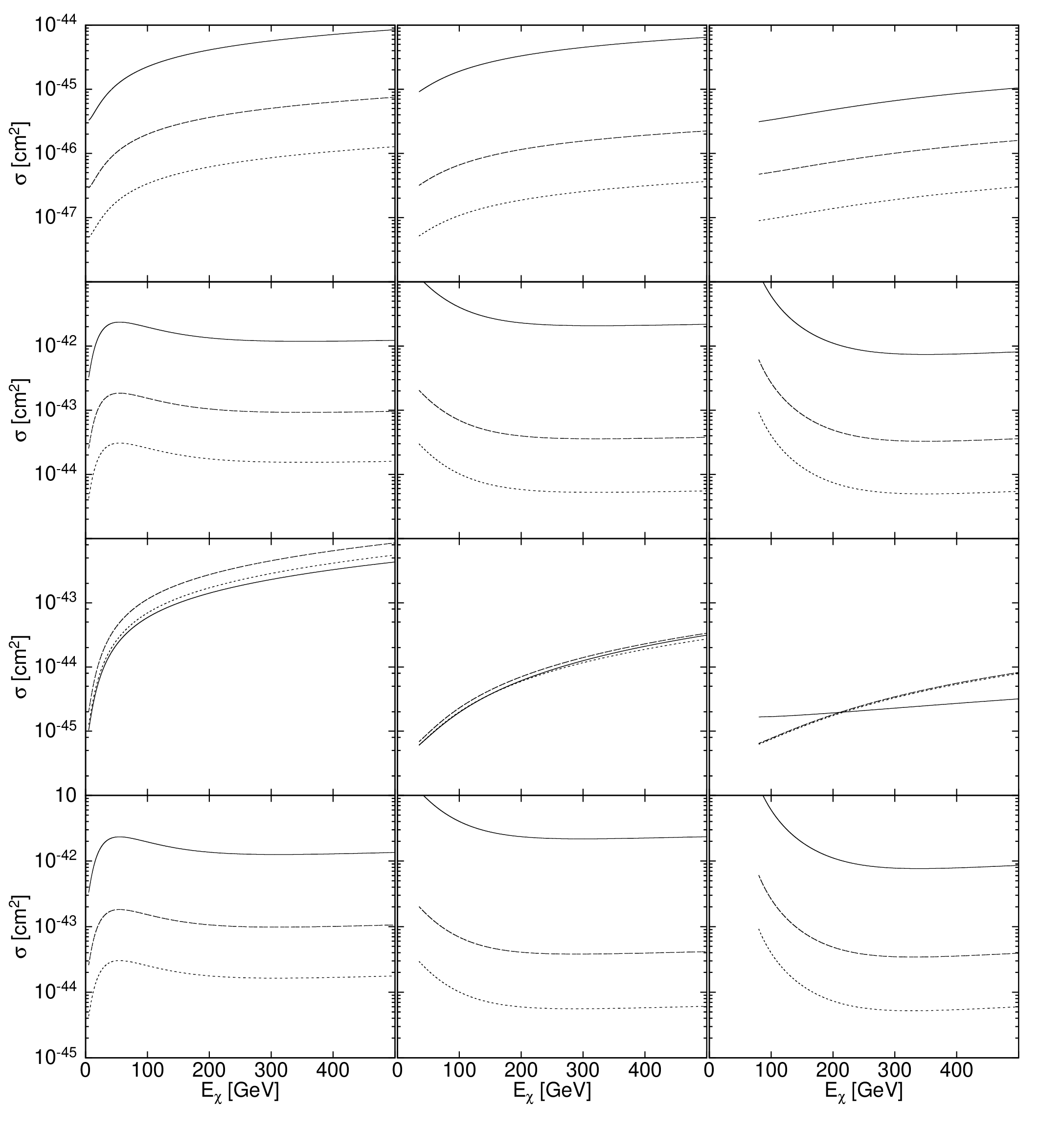, width=400pt, angle=-90}
\caption{Scalar contribution to the cross section as a function of the energy of the neutralino. Left column: $m_\chi=5$ GeV; middle column: $m_\chi=35$ GeV; right column: $m_\chi=80$ GeV. From top to bottom row: heaviest Higgs contribution; lightest Higgs contribution; squark contribution; total scalar contribution. Solid line: $\mu=100$ GeV; dashed line: $\mu=500$ GeV; dotted line: $\mu=1000$ GeV. The values for $M$ are listed in Table \ref{table-fig}.}
\label{sc-s-vs-e}
\end{figure*}

As for the case of the axial-vector terms we have studied the dependence of the cross section with the value of $\tan\beta$. Then, for e.g; $\tan \beta=10$, the resulting value of the scalar contribution to the cross section decreases by one order of magnitude. At quark level the interaction with the top-quark has the largest cross-section. The squark as a mediator produces the largest cross section for the up, down and strange quarks. If we compared our results with experimental data \cite{Ajaj:2019,abe19} we found some constraints on the value of $\mu$. For a neutralino mass $m_\chi=60$ GeV the higgsino parameter $\mu$ should be greater than $530$ GeV, and for $m_\chi=100$ GeV it should be $\mu>1370$ GeV.  For negative values of $\mu$ we have the following limits: $\mu<-193$ GeV for $m_\chi=60$ GeV, and $\mu<-372$ GeV for $m_\chi=100$ GeV.
\end{itemize}

\subsection{$m_\chi < 1$ GeV}

\begin{itemize}

\item Axial-vector contribution to the cross section

The results for the neutralino-nucleon cross section of the axial-vector channel, when the neutralino mass is of order of few MeV, are shown in Fig. \ref{av-s-vs-e-m-mev}.  The parameters used in the calculations are given in Table \ref{table-fig-mev}. The contribution due to the squark as a mediator is the same for protons or neutrons. However, if the mediator is a $Z_0$ boson, the neutralino-proton and neutralino-neutron cross sections are different, the neutron-neutralino cross section is larger than the proton-neutralino one. In both cases (protons and neutrons) the cross section decreases its value when the neutralino mass increases. The most important contribution to the axial-vector channel is given by the $Z_0$ boson as mediator.
\begin{table}[h!]
\begin{center}
\begin{tabular}{|c|c|c|}
\hline
$m_\chi$ [GeV] & $\mu$ [GeV] & $M$ [GeV] \\ \hline
$0.005$& $100$ & $61.36$ \\ \hline
$0.1$& $100$ & $61.59 $ \\ \hline
\end{tabular}
\end{center}
\caption{Used values of $m_\chi$, $\mu$ and $M$ in the computation of the cross section of Fig. \ref{av-s-vs-e-m-mev} and \ref{sc-s-vs-e-m-mev} for $\tan \beta=3$.}
\label{table-fig-mev}
\end{table}

\begin{figure}[h!]
\epsfig{file=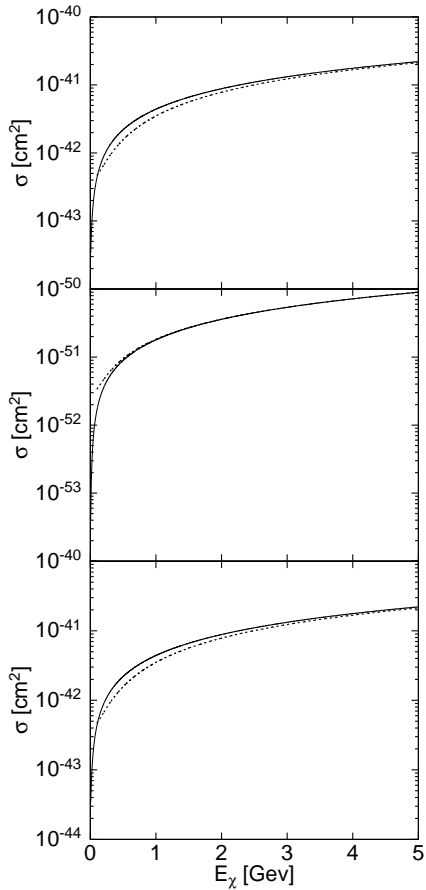, width=350pt, angle=-90}
\caption{Axial-vector contribution to the cross section as a function of the energy of the neutralino, for nucleon-light neutralino scattering. Top row: $Z_0$-boson contribution; middle row: squark contribution; bottom row: total axial-vector contribution. Solid line: $\mu=100$ GeV, $M=61.36$, and $m_\chi=5$ MeV, dotted line: $\mu=100$ GeV, $M=61.59$, and $m_\chi=100$ MeV.}
\label{av-s-vs-e-m-mev}
\end{figure}

If the value of $\tan\beta$ is modified (e.g. $\tan \beta=10$), the
contribution due to the squark as a mediator increases its value,
but it does not change the total axial-vector contribution. For
negative values of $\mu$ the cross section is much smaller that the
one obtained with $\mu>0$. If we considered the cross section
between the neutralino and quarks, the largest cross section
corresponds to the strange-quark and the smallest to the up-quark.

\item Scalar contribution to the cross section

In the case of the scalar channel the cross section for protons and neutrons are quite similar. In Fig. \ref{sc-s-vs-e-m-mev} we show the neutralino-proton cross section for this case. The three computed cross section for each one of the mediators, that is both Higgs and the squark, are of the same order of magnitude. If both Higgs are the mediator, the cross section increases its value when the neutralino mass is large, however, for the squark as a mediator the cross section is practically constant.
\begin{figure}[h!]
\epsfig{file=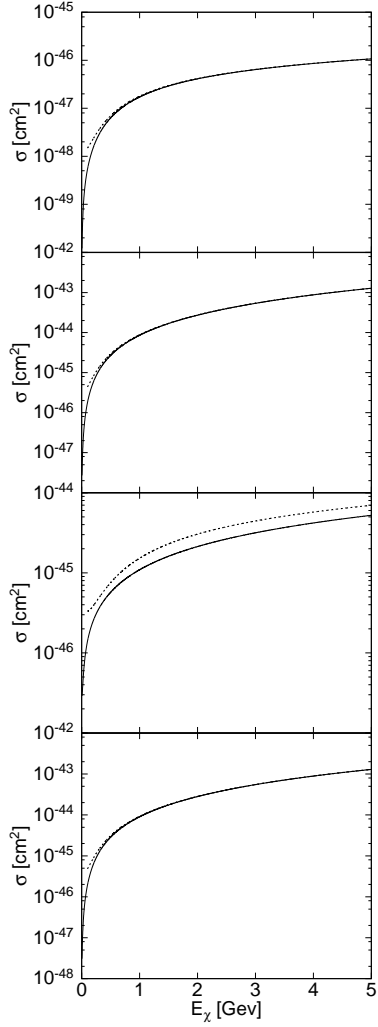, width=400pt, angle=-90}
\caption{Scalar contribution to the nucleon-light neutralino scattering cross section as a function of the energy of the neutralino for protons. From top to bottom row: heaviest Higgs, lightest Higgs, squark, and total scalar contributions, respectively. Solid line: $\mu=100$ GeV, $M=61.36$, and $m_\chi=5$ MeV, dotted line: $\mu=100$ GeV, $M=61.59$, and $m_\chi=100$ MeV.}
\label{sc-s-vs-e-m-mev}
\end{figure}

If $\tan\beta=10$ the total scalar contribution decreases its value one order of magnitude with respect to the previously presented case. Once again, the contribution due to the lightest (heaviest) Higgs as mediators decreases (increases) its value one order of magnitude. For negative values of $\mu$ the cross section is much smaller that the one for $\mu>0$. The interaction with the top-quark has the largest cross section while the smallest corresponds to the up-quark. The squark as a mediator produces the largest cross section for the up, down and strange quarks, while for the rest of the quarks the larger contribution to the cross section is given by the lightest Higgs.

\end{itemize}
%++++++++++++++++++++++++++++++++++++
\section{Conclusions}
\label{conclusion}

In this work we have studied the neutralino-nucleon cross section as a function of the neutralino mass and energy. We have considered different combinations of mass parameters of SUSY and different ratios between the vacuum expectation value of the Higgs. We have also analysed the quark-neutralino cross section for the different interactions, that is the axial-vector and scalar interactions.

We have found that the cross section for neutrons and protons are almost the same if the mass of the neutralino is larger than $1$ GeV. For smaller values of the neutralino mass the cross section, in the axial-vector channel, for neutrons is larger than the one for protons. The dominant mediator for the axial-vector channel is the $Z_0$ boson while for the scalar channel all the contributions are quite similar. If the value of $\mu$ increases the neutralino-nucleon cross section decreases.

In this work we have used $\tan \beta=3$ and $\tan \beta=10$ for the calculations of the cross section. We have found that the axial-vector contribution to the total cross section does not change its value when $\tan \beta$ changes, but the scalar contribution decreases its value for larger $\tan \beta$. We have also analysed the case with negative values of $\mu$ and found that the cross section is smaller than the one obtained with positive values of $\mu$. From the calculated values we could set constraints on the value of $\mu$ by the comparison with the available observational data.

%++++++++++++++++++++++++++++++++++++
\appendix

\section{Mass matrix diagonalization}
\label{diagonalizacion}

The mass-matrix in the framework of the MSSM is \cite{Elkheishen:1992}
\begin{eqnarray}
Y&=&\left(
\begin{array}{cccc}
M' & 0 & -M_Z c_{\beta} s_{W} & M_Z s_{\beta} s_W\\
0 & M & M_Z c_{\beta} c_W & - M_Z s_{\beta} c_W\\
-M_Z c_{\beta} s_W & M_Z c_{\beta} c_W & 0 & -\mu\\
M_Z s_{\beta}s_W & - M_Z s_{\beta} c_W & -\mu & 0\\
\end{array} \right)\, , \nonumber
\end{eqnarray}
where $c_{\beta}$ $\left(s_{\beta}\right)$ stands for $\cos{\beta}$ $\left(\sin{\beta}\right)$, $M_Z$ is the $Z$ boson mass, $\theta_W$ is the Weinberg angle and $c_W$ $\left(s_W\right)$ is $\cos{\theta_W}$ $\left(\sin{\theta_W}\right)$. The parameters $M$ and $M'$ are related by $M' = \frac{5}{3} M \tan^2{\theta_W}$ (GUT).

The lightest neutralino mass $\left(m_\chi\right)$ is the eigenvalue of the lowest mass-eigenstate of $Y$. Following Ref. \cite{Elkheishen:1992}, a squared, complex and unitary matrix $(N)$, (it can be real assuming CP invariance), transform $Y$ such that
\begin{eqnarray}
M_{diag}&=&N^{\dagger}YN \nonumber \\ &=&diag\left(\epsilon_1\,\tilde{M^0_1},\,\epsilon_2\,\tilde{M^0_2},\,\epsilon_3\,\tilde{M^0_3},\,\epsilon_4\,\tilde{M^0_4}\right)\, .
\end{eqnarray}
where $\epsilon_i=\pm 1$ and $M^0_i >0$. Taking into account the $i$-th row, one can find the following system of equations
\begin{eqnarray}
\left(
\begin{array}{cccc}
Y_{11}-\lambda_i & Y_{21} & Y_{31} & Y_{41}\\
Y_{12} & Y_{22}-\lambda_i & Y_{32} & Y_{42}\\
Y_{13} & Y_{23} & Y_{33}-\lambda_i & Y_{43}\\
Y_{14} & Y_{24} & Y_{34} & Y_{44}-\lambda_i\\
\end{array}
\right)\left(
\begin{array}{c}
Z_{i1} \\
Z_{i2} \\
Z_{i3} \\
Z_{i4} \\
\end{array}
\right)&=&0\, . \nonumber\\\label{ecua}
\end{eqnarray}
We have called $\lambda_i$ to the eigenvalue $\epsilon_i \,\tilde{M^0_i}$. After some algebra, the characteristic polynomial can be written as
\begin{eqnarray}
0 &=& \lambda_i^4- \zeta \, \lambda_i^3+ \xi \, \lambda_i^2+\gamma\, \lambda_i+\delta \, , \nonumber
\end{eqnarray}
with
\begin{eqnarray}
\zeta &=& M+M'\, ,\nonumber\\
\xi &=& MM'-\mu^2-M_Z^2\, ,\nonumber\\
\gamma &=& \left(M+M'\right)\mu^2+\left(M'\cos^2{\theta_W}+M\sin^2{\theta_W}\right)M_Z^2 \nonumber \\
&&-\mu M_Z^2\sin{2\beta}\, ,\nonumber\\
\delta &=& -MM'\mu^2+\left(M'\cos^2{\theta_W}+M\sin^2{\theta_W}\right)\mu M_Z^2\sin{2\beta}\, .\nonumber
\end{eqnarray}
The solutions of the characteristic polynomial are
\begin{eqnarray}
\left(\lambda_i\right)_1&=& \frac{\zeta}{4} -\sqrt{\frac{a}{4}-\frac{C_2}{6}-\frac{U}{12a}} \nonumber \\
&&+\sqrt{-\frac{a}{4}-\frac{C_2}{3}+\frac{C_3}{\sqrt{4a-\frac{8C_2}{3}-\frac{4U}{3a}}}+\frac{U}{12a}}\, ,\nonumber\\
\left(\lambda_i\right)_2&=& \frac{\zeta}{4}-\sqrt{\frac{a}{4}-\frac{C_2}{6}-\frac{U}{12a}} \nonumber \\
&&-\sqrt{-\frac{a}{4}-\frac{C_2}{3}+\frac{C_3}{\sqrt{4a-\frac{8C_2}{3}-\frac{4U}{3a}}}+\frac{U}{12a}}\, ,\nonumber\\
\left(\lambda_i\right)_3&=& \frac{\zeta}{4}+\sqrt{\frac{a}{4}-\frac{C_2}{6}-\frac{U}{12a}} \nonumber \\
&&-\sqrt{-\frac{a}{4}-\frac{C_2}{3}-\frac{C_3}{\sqrt{4a-\frac{8C_2}{3}-\frac{4U}{3a}}}+\frac{U}{12a}}\, ,\nonumber\\
\left(\lambda_i\right)_4&=& \frac{\zeta}{4}+\sqrt{\frac{a}{4}-\frac{C_2}{6}-\frac{U}{12a}} \nonumber \\
&&+\sqrt{-\frac{a}{4}-\frac{C_2}{3}-\frac{C_3}{\sqrt{4a-\frac{8C_2}{3}-\frac{4U}{3a}}}+\frac{U}{12a}}\, ,\nonumber
\end{eqnarray}
We have defined, following Ref.\cite{Elkheishen:1992},
\begin{eqnarray}
C_2&=&\xi-\frac{3}{8}\zeta^2 \, , \nonumber\\
C_3&=&-\frac{1}{8}\zeta^3+\frac{1}{2}\zeta \xi+\gamma \, , \nonumber\\
C_4&=&\delta+\frac{1}{4}\zeta\, \gamma+\frac{1}{16}\xi\, \zeta^2-\frac{3}{256}\zeta^4 \, , \nonumber\\
U&=&-\frac{1}{3}C_2^2-4C_4 \, , \nonumber\\
S&=&-C_3^2-\frac{2}{27}C_2^3+\frac{8}{3}C_2C_4 \, , \nonumber\\
D&=&-4U^3-27S^2 \, , \nonumber\\
a&=&\left(\frac{-S}{2}+\frac{1}{2}\sqrt{\frac{-D}{27}}\right)^{1/3} \, . \nonumber
\end{eqnarray}
To compute the eigenvectors we divide each equation of Eq. (\ref{ecua}) by $Z_{i1}$ (that is a not null value) and solve the system to obtain
\begin{eqnarray}
\frac{Z_{i2}}{Z_{i1}} &=& \frac{\lambda_i-M'}{\tan{\theta_W}\left(M-\lambda_i\right)} \, ,\nonumber\\
\frac{Z_{i3}}{Z_{i1}} &=& \frac{1}{2z} \left[-4\left(M-\lambda_i\right)\left(\lambda_i-M'\right)\mu \right. \nonumber\\
&&\left. \hskip 0.5cm+ M_Z^2 \left(-M+2\lambda_i -M'+\left(M-M'\right) \cos\left(2\theta_W\right)\right)\right. \nonumber\\
&&\left. \hskip 1.5cm \times \sin\left(2\beta\right)\right] \, , \nonumber\\
\frac{Z_{i4}}{Z_{i1}} &=& \frac{1}{z} \left[\left(\lambda_i-M'\right)\left(2\left(M-\lambda_i\right)\lambda_i +M_Z^2\left(1+\cos\left(2\beta\right)\right)\right) \right. \nonumber\\
&&\left. \hskip 0.5cm +2\left(M'-M\right) M_Z^2 \cos^2{\beta} \sin^2{\theta_W}\right]\, ,\nonumber
\end{eqnarray}
where
\begin{eqnarray}
z&=&2\sin{\theta_W}\left(M-\lambda_i\right)M_Z \left(\mu\cos{\beta}+\lambda_i\sin{\beta}\right)\, . \nonumber
\end{eqnarray}

Finally, using the normalization condition, that is
\begin{eqnarray}
Z_{i1}^2+Z_{i2}^2+Z_{i3}^2+Z_{i4}^2 &=& 1 \, , \nonumber
\end{eqnarray}
we obtain the eigenvector. We found that the eigenvalue $\left(\lambda_i\right)_1$ is positive and has the lowest value, therefore it could be associated to the neutralino mass.

%++++++++++++++++++++++++++++++++++++
\section{Matrix elements}
\subsection{Axial-Vector case}
\label{ap-AV}
The matrix element for the axial-vector interaction is
\begin{eqnarray}
M^{A-V}&=& \mathcal{C}^{A-V}\Big[\bar{\chi}\gamma^{\mu}\gamma_5\chi\Big]\Big[\bar{\psi_q}\gamma_{\mu}\gamma_5\psi_q\Big]\, .
\end{eqnarray}
Therefore one can split the expression into two contributions, one from the quark sector and the other one from the neutralino sector.
Since the quark current is $ J_{\mu}^{q}=\bar{\psi_q}\gamma_{\mu}\gamma_5\psi_q$, the quark contribution can be written as
\begin{eqnarray}
\sum_{t,t'}(J_{\mu}^{q} {J_{\alpha}^{q}}^{\dagger}) &=&\sum_{t,t'}\Big[\bar{\psi}_{q}(l',t') \gamma_{\mu}\gamma_5 {\psi}_{q}(l,t)\Big] \nonumber \\
&&\hskip 0.5cm \times \Big[\bar{\psi}_{q}(l',t')\gamma_{\alpha}\gamma_5 {\psi}_{q}(l,t)\Big]^{\dagger}\nonumber\\
&=& \tr\Big\{\gamma_{\mu}\gamma_5(\cancel{l}+m_{q}) \gamma_{\alpha}\gamma_5(\cancel{l'}+m_{q})\Big\}\label{av-quark}\\
&=& 4\left[\left(l_{\mu}l'_{\alpha}-g_{\mu\alpha}(l \cdot l')+l_{\alpha}l'_{\mu}\right)-m_{q}^2g_{\mu\alpha}\right]\, .\nonumber
\end{eqnarray}
For the neutralino sector, which current is $J^{\mu}_{\chi}=\bar{\chi}\gamma^{\mu}\gamma^5\chi$, we obtain
\begin{eqnarray}
\sum_{s,s'}(J^{\mu}_{\chi} {J^{\alpha}_{\chi}}^{\dagger}) &=&\sum_{s,s'}\Big[\bar{\chi}(p',s')\gamma^{\mu}\gamma^5 {\chi}(p,s)\Big] \nonumber \\
&&\hskip 0.5cm \times \Big[\bar{\chi}(p',s')\gamma^{\alpha}\gamma^5{\chi}(p,s)\Big]^{\dagger}
\label{av-neutralino}\\
 &=&4\left[\left(p^{\mu}p'^{\alpha}-g^{\mu\alpha}(p \cdot p')+p^{\alpha}p'^{\mu}\right)-m_{\chi}^2g^{\mu\alpha}\right]\, . \nonumber
\end{eqnarray}
Finally the squared invariant matrix element is
\begin{eqnarray}
\sum_{s,s',t,t'}\vert M^{A-V}\vert^2 &=&32 \left|\mathcal{C}^{A-V}\right|^2 \left[(p \cdot l)(p'\cdot l')+(p\cdot l')(p'\cdot l)\right. \nonumber \\
&&\left. \hskip 2cm +m_{\chi}^2(l\cdot l')+m_q^2(p\cdot p')\right. \nonumber \\
&&\left. \hskip 2cm  +2m_q^2m_{\chi}^2\right]\, .  \label{M-AV}
\end{eqnarray}

\subsection{Scalar case}
\label{ap-SC}

The matrix element for the scalar interaction is
\begin{eqnarray}
M^{S}&=& \mathcal{C}^{S}\Big[\bar{\chi}\chi\Big]\Big[\bar{\psi_q}\psi_q\Big]\, ,
\end{eqnarray}
Once again, one can split the expression into two contributions, one from the quark sector and the other one from the neutralino sector. For the quark sector, we derive
\begin{eqnarray}
 \sum_{t,t'}(J^{q} {J^{q}}^{\dagger}) &=&\sum_{t,t'}\Big[\bar{\psi}_{q}(l',t') {\psi}_{q}(l,t)\Big] \Big[\bar{\psi}_{q}(l',t') {\psi}_{q}(l,t) \Big]^{\dagger} \nonumber\\
 &=& 4(l\cdot l'+m_{q}^2)\, , \label{sc-quark}
\end{eqnarray}

The neutralino current is $ J_{\chi}=\bar{\chi}\chi$, therefore
\begin{eqnarray}
 \sum_{s,s'}(J_{\chi} {J_{\chi}}^{\dagger}) &=&\sum_{s,s'}\left[\bar{\chi}(p',s') {\chi}(p,s)\right] \left[\bar{\chi}(p',s') \chi(p,s) \right]^{\dagger} \nonumber\\
 &=&4(p\cdot p'+m_{\chi}^2) \, ,
 \label{sc-neutr}
\end{eqnarray}

Finally the squared invariant matrix element is written
\begin{eqnarray}
\sum_{s,s',t,t'}\vert M^{S}\vert^2 &=& 16\left|\mathcal{C}^{S}\right|^2 \left[(p \cdot p')(l\cdot l') +m_{\chi}^2(l\cdot l')\right. \nonumber \\
&&\left. \hskip 1.5cm +m_q^2(p\cdot p')+m_q^2m_{\chi}^2\right]\, .
\label{M-esc}
\end{eqnarray}

\section{Acknowledgments}
This work was supported by the grant PIP-616 of the National Research Council of Argentina (CONICET), and by a research-grant of the National Agency for the Promotion of Science and Technology (ANPCYT) of Argentina. O. C. and M. E. M. are members of the Scientific Research Career of the CONICET. K. F. is a PhD. fellow of CONICET.

%\bibliographystyle{ieeetr}
%\bibliographystyle{apsrev}
%\bibliography{referencias-prc}

\begin{thebibliography}{99}
\bibitem{Zwicky:1933} F. Zwicky, Helv. Phys. Acta {\bf{6}}, 110 (1933).
\bibitem{Rubin:1970}  V. C. Rubin and J. Ford, W. Kent, Astrophysical Journal {\bf{159}}, 379 (1970).
\bibitem{Gelmini:2017} G. B. Gelmini, Rept. Prog. Phys. {\bf{80}}, 082201 (2017).
\bibitem{Roszkowski:2018}   L. Roszkowski, E. M. Sessolo, and S. Trojanowski, Reports on Progress in Physics {\bf{ 81}}, 066201 (2018).
\bibitem{majumdar:2014} D. Majumdar, Dark Matter: An Introduction (Taylor and Francis, 2014), ISBN 9781466572119.
\bibitem{Chung:2001} D. J. H. Chung, P. Crotty, E. W. Kolb, and A. Riotto, Phys. Rev. D 64, 043503 (2001).
\bibitem{Kolb:1999} E. W. Kolb, D. J. H. Chung, and A. Riotto, in Dark matter in Astrophysics and Particle Physics, edited by H. V.
Klapdor-Kleingrothaus and L. Baudis (1999), p. 592, hep-ph/9810361.
\bibitem{Alcantara:2019} E. Alcantara, L. A. Anchordoqui, and J. F. Soriano, Phys. Rev. D 99, 103016 (2019).
\bibitem{Undagoitia:2015} T. M. Undagoitia and L. Rauch, Journal of Physics G: Nuclear and Particle Physics 43, 013001 (2015).
\bibitem{Paczynski:1986} B. Paczynski, Astrop. J. 304, 1 (1986).
\bibitem{Alcock:2000} C. Alcock, R. A. Allsman, D. R. Alves, T. S. Axelrod, A. C. Becker, D. P. Bennett, K. H. Cook, N. Dalal, A. J. Drake,
K. C. Freeman, et al., The Astrophysical Journal 542, 281 (2000).
\bibitem{Planck:2018} N. Aghanim et al. (Planck), ArXiv e-prints (2018), 1807.06209.
\bibitem{Perez:2020} P. F. Perez, C. Murgui, and A. D. Plascencia, Journal of High Energy Physics 2020, 91 (2020).
\bibitem{Freese:2017} K. Freese, International Journal of Modern Physics D 26, 1730012-223 (2017).
\bibitem{Dodelson:1994} S. Dodelson and L. M. Widrow, Phys. Rev. Let. 72, 17 (1994).
\bibitem{Boyarsky:2019} A. Boyarsky, M. Drewes, T. Lasserre, S. Mertens, and O. Ruchayskiy, Progress in Particle and Nuclear Physics 104, 1
(2019).
\bibitem{Campos:2016} M. D. Campos and W. Rodejohann, Phys. Rev. D 94, 095010 (2016).
\bibitem{Divari:2017} P. C. Divari and J. D. Vergados, arXiv e-prints 1707.02550 (2017).
\bibitem{Freese:2012} K. Freese, M. Lisanti, and C. Savage, ArXiv e-prints (2012), 1209.3339.
\bibitem{davis:2015} J. H. Davis, Int. J. Mod. Phys. A30, 1530038 (2015).
\bibitem{Scopel:2008} S. Scopel, in Journal of Physics Conference Series (2008), vol. 120 of Journal of Physics Conference Series, p. 042003.
\bibitem{Gelmini:2016} G. B. Gelmini, in Journeys Through the Precision Frontier: Amplitudes for Colliders (TASI 2014), edited by L. Dixon
et al. (2016), pp. 559.
\bibitem{xenon}E. Aprile et al. [XENON], [arXiv:2006.09721 [hep-ex]].
\bibitem{lee} H-M.Lee,[ arXiv:2006.13183v1[hep-ph]].
\bibitem{Goodman:1985} M. W. Goodman and E. Witten, Phys. Rev. D 31, 3059 (1985).
\bibitem{Drukier:1986} A. K. Drukier, K. Freese, and D. N. Spergel, Phys. Rev. D 33, 3495 (1986).
\bibitem{Conrad:2014} J. Conrad, arXiv e-prints 1411.1925 (2014).
\bibitem{Bernabei:2018} R. Bernabei et al., Universe 4, 116 (2018).
\bibitem{Amare:2019} J. Amare et al., Phys. Rev. Lett. 123, 031301 (2019).
\bibitem{cdex:2018} H. Jiang et al. (CDEX Collaboration), Phys. Rev. Lett. 120, 241301 (2018).
\bibitem{Abdelhameed:2019} A. H. Abdelhameed et al. (CRESST) (2019), 1904.00498.
\bibitem{CDMS:2016} R. Agnese et al. (SuperCDMS Collaboration), Phys. Rev. Lett. 116, 071301 (2016).
\bibitem{XENON1T:2018} E. Aprile et al. (XENON Collaboration 7), Phys. Rev. Lett. 121, 111302 (2018).
\bibitem{Antonello:2018} M. Antonello et al. (SABRE), Eur. Phys. J. C 79, 363 (2019).
\bibitem{Santos:2013} D. Santos, G. Bosson, J. L. Bouly, O. Bourrion, et al., Journal of Physics: Conference Series 469, 012002 (2013).
\bibitem{Adhikari:2019} G. Adhikari et al. (COSINE-100), Phys. Rev. Lett. 123, 031302 (2019).
\bibitem{DarkSide-50:2018} P. Agnes et al. (DarkSide Collaboration), Phys. Rev. Lett. 121, 081307 (2018).
\bibitem{Amole:2019} C. Amole et al. (PICO), Phys. Rev. D 100, 022001 (2019).
\bibitem{Ajaj:2019} R. Ajaj et al. (DEAP), Phys. Rev. D 100, 022004 (2019).
\bibitem{ANDES_LAB} O. Civitarese, Nuclear and Particle Physics Proceedings 267-269, 377 (2015).
\bibitem{Elkheishen:1992} M. M. El Kheishen, A. A. Shafik, and A. A. Aboshousha, Physical Review D 45, 4345 (1992).
\bibitem{bertone04} G. Bertone, D. Hooper, and J. Silk, Phys. Rep. 405, 279 (2004).
\bibitem{Murakami:2001} B. Murakami and J. D. Wells, Phys. Rev. D 64, 015001 (2001).
\bibitem{Ellis:2000} J. Ellis, A. Ferstl, and K. A. Olive, Physics Letters B 481, 304 (2000).
\bibitem{Cerdeno:2001} D. G. Cerdeno, S. Khalil, and C. Munoz, arXiv e-prints hep-ph/0105180 (2001).
\bibitem{Engel:1992} J. Engel, S. Pittel, and P. Vogel, International Journal of Modern Physics E 1, 1 (1992).
\bibitem{Djouadi:2008} A. Djouadi, Physics Reports 459, 1 (2008), ISSN 0370-1573.
\bibitem{Tanabashi:2018} M. Tanabashi et al. (Particle Data Group), Phys. Rev. D 98, 030001 (2018).
\bibitem{Gondolo:2004} P. Gondolo, J. Edsjö, P. Ullio, L. Bergström, M. Schelke, and E. A. Baltz, Journal of Cosmology and Astrophysics 2004,
008 (2004).
\bibitem{Lisanti:2016} M. Lisanti, in Proceedings, Theoretical Advanced Study Institute in Elementary Particle Physics: New Frontiers in Fields
and Strings (TASI 2015): Boulder, CO, USA, June 1-26, 2015 (2017), pp. 399–446, arXiv e-prints 1603.03797.
\bibitem{behnke17} E. Behnke, M. Besnier, P. Bhattacharjee, X. Dai, et al., Astroparticle Physics 90, 85 (2017).
\bibitem{xia19} J. Xia, A. Abdukerim, W. Chen, X. Chen, et al., Physics Letters B 792, 193 (2019), ISSN 0370-2693.
\bibitem{suzuki19} Xmass Collaboration, T. Suzuki, K. Abe, K. Hiraide, K. Ichimura, et al., Astroparticle Physics 110, 1 (2019).
\bibitem{abe19} Xmass Collaboration, K. Abe, K. Hiraide, K. Ichimura, Y. Kishimoto, et al., Physics Letters B 789, 45 (2019).
\end{thebibliography}

\end{document}